\documentclass[fleqn,usenatbib]{mnras}

\usepackage{newtxtext,newtxmath}

\usepackage[T1]{fontenc}

\DeclareRobustCommand{\VAN}[3]{#2}
\let\VANthebibliography\thebibliography
\def\thebibliography{\DeclareRobustCommand{\VAN}[3]{##3}\VANthebibliography}

\makeatletter
\def\endfigure{\end@float}
\makeatother

\makeatletter
\def\endtable{\end@float}
\makeatother

\usepackage[utf8]{inputenc}
\usepackage{graphicx}
\usepackage{amsmath}

\usepackage{amssymb}
\usepackage{booktabs}
\usepackage{array}
\usepackage{tabularx}
\usepackage{longtable}
\usepackage{float}
\usepackage{placeins}
\usepackage{geometry}
\usepackage{makecell}
\usepackage{adjustbox}
\usepackage{threeparttable}
\usepackage{tabularray}
\usepackage{nicematrix}
\usepackage{multicol}
\usepackage{multirow}
\usepackage{caption}
\usepackage{subcaption}
\usepackage{cleveref}
\usepackage{hyperref}
\usepackage{enumitem}

\title[X-ray sources in NGC 6528]{The X-ray Source Population of the Metal-Rich Globular Cluster NGC 6528}

\author[B. Leal et al.]{
Bernard Leal$^{1}$,
Kwangmin Oh$^{1}$ \thanks{E-mail: min8582046@gmail.com},
Jay Strader$^{1}$,
Steve E. Zepf$^{1}$,
Kristen Dage$^{2}$,
S. Kim$^{3}$,
C.Y. Hui$^{4}$
\\
$^{1}$Michigan State University, Department of Physics and Astronomy, Michigan State University, East Lansing,  MI 48824, USA\\
$^{2}$International Centre for Radio Astronomy Research -- Curtin University, GPO Box U1987, Perth, WA 6845, Australia\\
$^{3}$Department of Earth, Environmental \& Space Sciences, Chungnam National University, Daejeon 34134, Republic of Korea\\
$^{4}$Department of Astronomy and Space Science, Chungnam National University, Daejeon 34134, Republic of Korea
}

\date{Accepted XXX. Received YYY; in original form ZZZ}

\pubyear{\the\year{}}

\begin{document}
\label{firstpage}
\pagerange{\pageref{firstpage}--\pageref{lastpage}}
\maketitle

\begin{abstract}
We present the first study of the X-ray sources in one of the most metal-rich globular clusters in the Galaxy, NGC 6528. Using relatively deep (66 ksec) archival imaging from the {\it Chandra} X-ray Observatory, we identify 18 sources within the half-light radius of the cluster, all in the range $L_X \sim 10^{31}$--$10^{32}$~erg~s$^{-1}$ (0.5--7 keV). By combining these data with photometry from the Hubble Space Telescope and other sources, we classify the X-ray sources as a likely mix of cataclysmic variables and active binaries, though one or more of the brighter objects could be a quiescent low-mass X-ray binary. For this cluster, it appears that the X-ray binary-enhancing effects of high metallicity are outweighed by the cluster's advanced dynamical evolution, leading to a relatively modest X-ray source population.
\end{abstract}

\begin{keywords}
stars: activity -- binaries: close -- globular clusters: individual: NGC 6528 -- X-rays: binaries
\end{keywords}



\section{Introduction}
Globular clusters (GCs) are one of the densest stellar systems, especially in their central regions. Due to the frequent stellar dynamical interactions, they efficiently produce exotic objects such as low-mass X-ray binaries (LMXBs), and millisecond pulsars (MSPs), and can also dynamically form cataclysmic variables (CVs) and active binaries (ABs). Consequently, GCs host a significantly higher number of X-ray sources per unit mass compared to the field, providing key insights into the influence of the dense stellar environments that host various close binary systems \citep[e.g.][]{clark_1975, hills_1976, Heinke_2005, Ivanova_2006, Ivanova_2008}. 

The role of binaries and their hosting GC environments has been extensively studied through observations and simulations. Using the exceptional spatial resolution of the \textit{Chandra} X-ray Observatory \citep{Weisskopf_2000}, diverse populations of X-ray-emitting binaries have been uncovered in numerous GCs, shedding light on their formation and evolutionary processes \citep[e.g.,][]{Bassa_2004, Heinke_2005, Henleywillis2018MNRAS.479, Oh2020.498, Bahramian_2020,zhao_2020}. Furthermore, compelling evidence for the influence of dynamics in forming such binary systems has been found across different clusters, emphasizing the importance of stellar encounters in their formation \citep[e.g.,][]{Pooley_2003, Ivanova_2005, Pooley_Hut_2006, fregeau_2009, Hui_2010, Oh_2024}.

Numerical studies from Monte Carlo simulations, such as the Monte Carlo cluster simulator (MOCCA) \citep{Giersz_2013} and Cluster Monte Carlo code (CMC) \citep{Kremer_2020,Rodriguez_2022}, have also advanced our understanding of how the dynamical evolution of GCs can influence their compact binary populations. In clusters with high central densities or those that have undergone core collapse, the increased frequency of dynamical interactions promotes both the formation and hardening of compact binaries. These processes result in tighter binary orbits and more efficient mass transfer, leading to higher accretion rates and enhanced X-ray luminosities. Studies focusing on CVs indicate that more dynamically evolved clusters tend to host brighter X-ray sources, suggesting that stellar encounters can influence the accretion properties of binaries \citep[e.g.,][]{Hong_2017, belloni4, Oh_2024}. 

Previous studies have highlighted the role of metallicity in shaping the X-ray properties of binaries. Higher-metallicity GCs tend to host more X-ray binaries per unit mass, as observed across different environments and luminosity ranges \citep{Kundu2003, Kim_2013, Vulic_2018, Heinke_2020}. While the underlying mechanisms remain under investigation, metallicity may influence binary evolution by modifying donor star properties, altering opacities, and affecting accretion physics, thereby impacting both the formation efficiency and luminosity distribution of X-ray binaries.

Within this context, NGC 6528 represents a particularly intriguing target for studying the interplay between metal-rich environments and the formation of compact binaries. NGC 6528 is one of the most metal-rich GCs, with [Fe/H] $\approx -0.1$ \citep{Ortolani_2001,Zoccali_2004,Lagioia_2014,munoz_2018}, and is located in Baade’s Window near the Galactic Centre (0.6 kpc separation; \citealp{Ortolani_1995,Barbuy_1998,Harris_2010}). 

In this paper, we use archival data from the \textit{Chandra} X-ray Observatory to identify 18 X-ray sources within the half-light radius of NGC 6528.  Our X-ray analysis focuses on variability, spectral properties, and flux measurements. We supplement the X-ray data with
optical observations from the Hubble space telescope (\textit{HST}), \textit{Gaia}, and other complementary photometric surveys to identify potential optical counterparts to the X-ray sources. Together, these data allow us to characterize the binaries and explore how factors such as metallicity and the cluster’s dynamical environment influence their evolution. 

\section{\textit{Chandra} X-ray Observatory}

\subsection{Data reduction}
We analysed two archival observations of NGC 6528 obtained with the Advanced CCD Imaging Spectrometer (ACIS) onboard the \textit{Chandra} X-ray Observatory. The deeper observation was made on 9 February 2011 with an exposure of 54.3 ks (Obs ID 12400, PI Kong, referred to as obs1), and the shallower observation on 5 May 2008 with an exposure of 12.2 ks (Obs ID 8961, PI Pooley, referred to as obs2). A summary of these observations is provided in \autoref{tab:obs_info}. In both cases, the cluster centre was placed on ACIS-S on the sensitive S3 chip.

\begin{table}
        \caption{\textit{Chandra} Observations on NGC 6528}
        \label{tab:obs_info}
        \renewcommand{\arraystretch}{1.3}
        \begin{adjustbox}{width=0.47\textwidth}
        \begin{tabular}{cccc}
 \hline
{Telescope/Instrument} & {Observation Date} & {Observation ID} & {Exposure Time 
(ks)} \\
\hline
$Chandra$/ACIS-S & 2011 Feb 09 & 12400 & 54.3 \\
$Chandra$/ACIS-S & 2008 May 05 & 8961 & 12.2 \\
\hline
\end{tabular}
\end{adjustbox}
\end{table}

Using \texttt{CIAO} (version 4.17), we reprocessed both data with the \texttt{chandra\_repro} script and merged the observations with \texttt{merge\_obs}, using broadband, bin=1 (native resolution of ACIS), and per-observation PSF maps using an encircled-energy fraction (psfecf) of 0.393, as well as a combined PSF map, weighting the ECF values by the observation exposure times (psfmerge=exptime). The analysis was restricted to an energy band of 0.5--7.0 keV. 

\subsection{Source detection}
We used the \texttt{CIAO} task \texttt{wavdetect}, which identifies sources by applying wavelet transformations at multiple spatial scales. This method is particularly well-suited for crowded fields, such as the dense stellar regions in GCs, since it can separate closely spaced point sources as well as find extended sources. For this analysis, we fixed pixel scales at 0.5, 1.0, 1.4, 2.0, 4.0, and 8.0 to capture sources of varying angular sizes. A detection threshold of $10^{-6}$ was adopted, corresponding to a false positive rate of approximately one spurious source per 10$^6$ pixels. The maximum number of iterations was set to ten to ensure the convergence of source boundaries. In this work, only sources with a signal-to-noise ratio (S/N) > 3 were considered genuine, ensuring robust detections and minimizing contamination from noise or spurious artifacts.

\begin{figure}
    \centering
    \includegraphics[width=\linewidth]{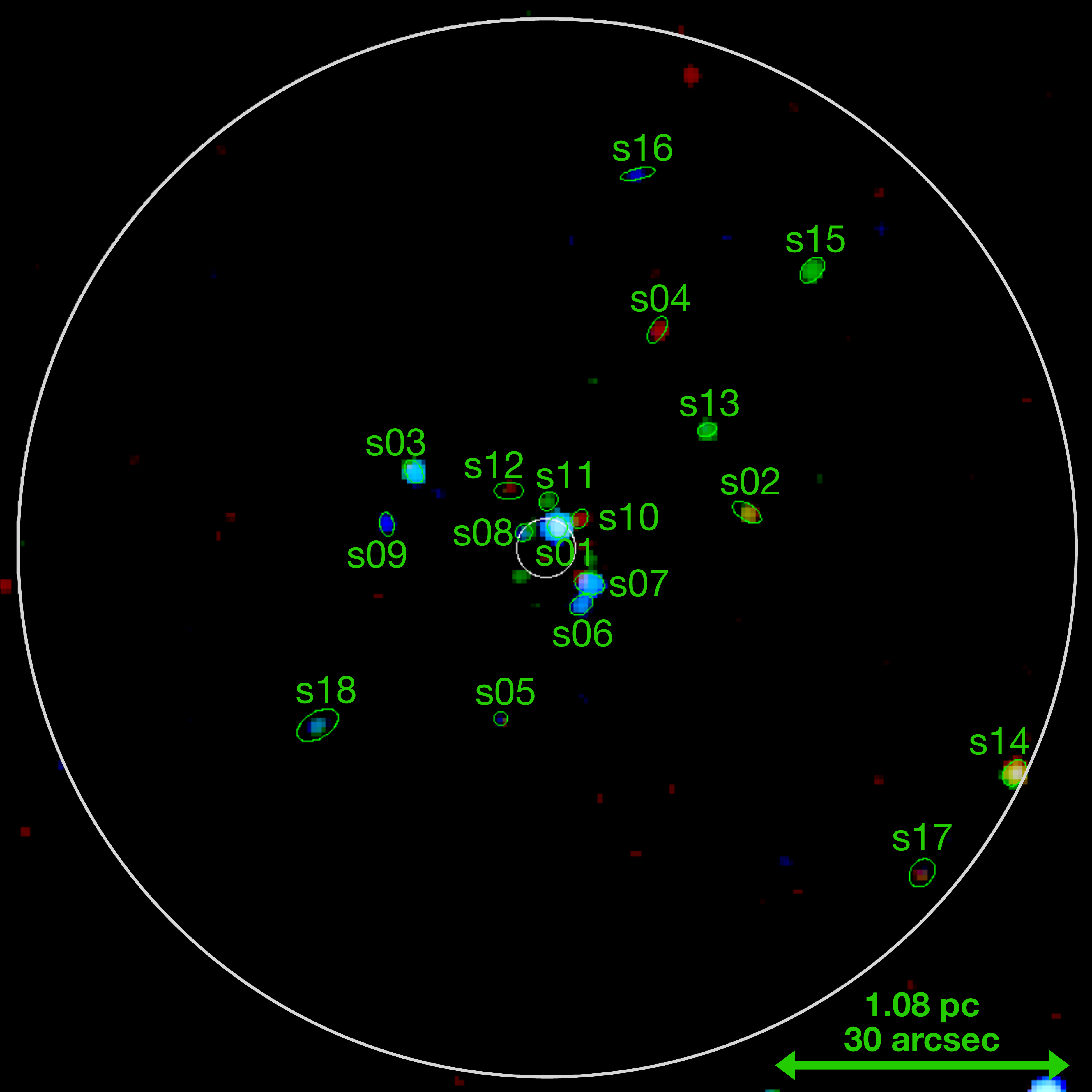}
    \caption{\textit{Chandra} Energy-coded X-ray image of NGC 6528 with merged of two \textit{Chandra} ACIS observations. The outer and inner circles represent the half-light radius (0.91 arcmin) and the core radius (3.05 arcsec), respectively. The colour image consists of red (soft: 0.5 - 1.0 keV), green (medium: 1.0 - 2.0 keV), and blue (hard: 2.0 - 7.0 keV). A total of 18 sources > 3$\sigma$ were detected, represented by the green circles.}
    \label{rgb_img}
\end{figure}

A total of 18 X-ray sources were detected within the half-light radius (0.91 arcmin; \citealt{Baumgardt_2018}) of NGC 6528. The properties of these sources, including their positions, fluxes, and variability characteristics, are summarized in \autoref{tab:results}. \autoref{rgb_img} presents a colour-coded X-ray image of the cluster, with red (0.5–1.0 keV), green (1.0–2.0 keV), and blue (2.0–7.0 keV) representing the energy bands. The detected sources are marked with green ellipses and labeled according to the numbering in \autoref{tab:results}. Given that the core radius of the cluster is extremely small ($\sim$0.05 arcmin), we do not present a separate core region analysis. However, we note that sources s01 and s08 are located within the core radius.

We attempted to improve the absolute astrometry of the Chandra data by matching X-ray sources outside the cluster half-light radius to potential Gaia counterparts. However, owing to the cluster's location deep within the Bulge, the Gaia source density is extraordinarily high, which makes secure cross-matching unreliable. Hence, we were not able to refine the absolute astrometry in this manner.

We also examined the flux values of these sources in the individual observations to assess their variability. Fluxes were calculated using the \texttt{srcflux} script (Section 3.1). To maximize the detection probability, we used the source regions obtained from the merged observation. Five sources (s03, s04, s06, s11, and s16) exhibited no measurable flux in obs2. Therefore, we used the upper limit values of those sources in the \autoref{tab:results}.

Here, we classified them as long-term variable candidates if the source is detected as S/N larger than 3 in the merged observation but yields no measurable flux in the shorter observation. These sources are flagged with L$^*$ in the variability column of \autoref{tab:results}. The remaining sources were detected in both observations, with varying flux levels, which are further discussed in Section 3.2.

\section{X-ray data analysis}

\subsection{X-ray Colour-Luminosity Diagram}
We estimated the unabsorbed energy flux of all the sources in the merged observation using the \texttt{srcflux} script. The analysis was performed uniformly, assuming a power-law model with a photon index = 2 (in Section 3.3 below, we relax this assumption for the brightest sources with sufficient counts to constrain their spectra). In these fits, the hydrogen column density ($N_H$) was fixed at 3.7 $\times$ 10 $^{21}$ cm$^{-2}$, consistent with the foreground extinction toward NGC 6528. This value was derived using the optical extinction \(E(B-V) = 0.54\) reported by \cite{Harris_2010}. Subsequently, optical extinction ($A_V$) was converted to $N_H$ as described by \cite{Guver2009.400}. This conversion is appropriate for the elemental abundances adopted by \citet{Anders_grevesse_1989}, which were used throughout our X-ray spectral analysis. We fit the model three separate times, each considering a credible interval of 1$\sigma$ (68\% confidence level), for three different energy bands: a broadband (0.5--7.0 keV), a soft+medium band (0.5--2.0 keV), and a hard band (2.0--7.0 keV).

The X-ray luminosities of the sources were calculated by assuming a distance of 7.9 kpc to NGC 6528 from \cite{Harris_2010}. For the X-ray colour, we adopted the logarithmic ratio of counts in the soft+medium and hard bands, defined as 2.5log((S+M)/H), where S+M and H are the net counts in the soft+medium band (0.5--2.0 keV) and hard band (2.0--7.0 keV), respectively.

\begin{figure} 
    \centering 
    \includegraphics[width=\linewidth]{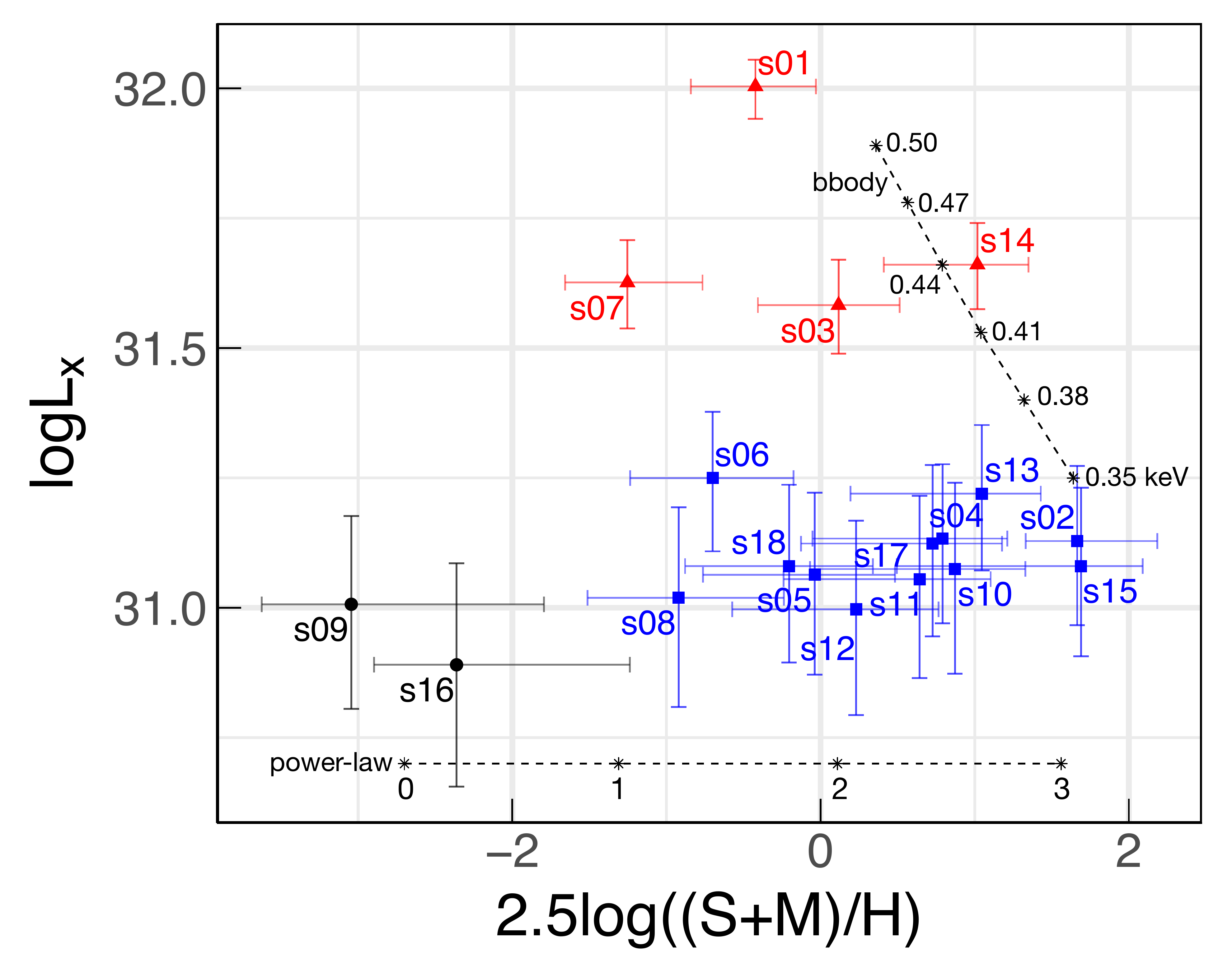}
    \caption{X-ray colour-luminosity diagram of 18 detected X-ray sources within the half-light radius of NGC 6528. The symbol colours (black, blue, and red) represent the classification results from the Gaussian Mixture Model. We note that these colours do not necessarily correspond to specific source types. The plotted error bars show 1$\sigma$ uncertainties of both colour and luminosity. Comparison spectral models (power-law or blackbody) are shown as dotted lines. For \textit{bbody} model, an emitting region radius of 0.1km is assumed.}
    \label{fig:xray_CMD}
\end{figure}

\autoref{fig:xray_CMD} shows the X-ray colour magnitude diagram (CMD) for the 18 detected sources, with the luminosity plotted on the y-axis and the X-ray colour on the x-axis. The error bars represent 1$\sigma$ uncertainties in both luminosity and colour. A Gaussian Mixture Model was applied to identify clusters of sources based on their positions in the diagram. We employed the CRAN Mclust package (version 5.4.6; \cite{Scrucca_2016}) to fit the models and calculate the likelihoods for each. The selection of the optimal model was guided by the Bayesian information criterion \citep{Schwarz_1978}. The clusters are colour-coded in \autoref{fig:xray_CMD} to reflect the mixture model classification. This figure also shows representative power-law and blackbody models for comparison to the measurements.

\subsection{Variability analysis} 
The availability of two \textit{Chandra} observations of NGC 6528, obtained approximately three years apart, allows us to investigate both long-term and short-term X-ray flux variability. Long-term variability was assessed by comparing fluxes between the two epochs, while short-term variability was analysed within each individual observation.

\subsubsection{Long-term variability}
To quantify the long-term X-ray flux variability of the sources, we computed the $S_{\text{flux}}$ by the following equation:
\[
S_{\text{flux}} = \frac{\left| F_{\text{obs1}} - F_{\text{obs2}} \right|}{\sqrt{\sigma^2_{\text{obs1}} + \sigma^2_{\text{obs2}}}} \tag{1}
\]
where ${F_{\text{obs1}}}$ and ${F_{\text{obs2}}}$ are the absorption-corrected X-ray fluxes from the first and second observations, respectively, and ${\sigma^2_{\text{obs1}}}$ and ${\sigma^2_{\text{obs2}}}$ are the uncertainties in these fluxes. For the error terms, we used the average of the upper and lower bounds provided by \texttt{srcflux}.

Hence, $S_{\text{flux}}$ indicates the significance of the flux variation between two epochs. In this work, we consider an X-ray source to exhibit long-term variability if its $S_{\text{flux}}$ is larger than 3, as this corresponds to a 3$\sigma$ significance level, ensuring the variation is unlikely to be due to random fluctuations. Using this criterion, s01 was identified as a long-term variable source, as indicated by the "L" designation in \autoref{tab:results}. 

In contrast, several sources (e.g., s03, s04, s06, s11, and s16) show significant flux in the first (54 ks) observation but are not detected in the second (12 ks) observation. For these sources, we report the 90\% confidence upper limits derived from \texttt{srcflux} in the second epoch in \autoref{tab:results}. Among them, if the flux difference remains significant even when using the upper limit in the second epoch—i.e., exceeding the 3$\sigma$ threshold—we assign an "L*" flag instead of "L" to indicate long-term variability under upper-limit constraints. Further confirmation through deeper observations or statistical modeling would help verify their variability more robustly.

The long-term variability observed in these sources could result from changes in accretion rates, eclipses in interacting binaries, or other processes typical of GC X-ray sources.

\subsubsection{Short-term variability}
We used the \texttt{CIAO} script \texttt{glvary} that is based on the Gregory-Loredo algorithm to search for the short-term variability in each observation \citep{Gregory.Loredo1992.398}. The algorithm breaks a single observation into multiple time bins and searches for any deviations between different bins. The variability is ranked by an index with a range from 0 to 10. In our work, we consider a source to be variable in a particular observation if its variability index computed by glvary is $\geq$6, corresponding to a probability of $\geq$ 90 $\%$ that the signal is genuinely variable. Applying this criterion, we found no evidence of short-term X-ray flux variability in either of the observations.

\begin{table*}
	\centering
        \caption{X-ray properties of X-ray sources within the half-light radius of NGC 6528.}
        \label{tab:results}
        \renewcommand{\arraystretch}{1.3}
        \setlength{\tabcolsep}{3.1pt}
        \begin{tabular}{ccccccccccccccccc}
 \hline
{Source} & {RA}$^{(1)}$ & $\delta_{RA}$ & {Dec.}$^{(1)}$ & $\delta_{Dec}$& {S/N}$^{(2)}$ & {$F_{\rm obs1}$}$^{(3)}$ & {$F_{\rm obs2}$}$^{(3)}$ & {$F_{\rm merged}$}$^{(4)}$ & {$F_{\rm S+M}$}$^{(5)}$ & {$F_{\rm H}$}$^{(5)}$ & {$L_x$}$^{(6)}$ & {Variability}$^{(7)}$
\\
 &  
 &
 \scriptsize(arcsec) &
 &  
 \scriptsize(arcsec) &  
 &
 \multicolumn{5}{c}{($10^{-15}$~erg~cm$^{-2}$~s$^{-1}$)}
 &
 \scriptsize{($10^{31}$~erg~s$^{-1}$)} & 
 \scriptsize{(L,L*)}
\\

\hline
\noalign{\vspace{0.1cm}}
$s01$ & 18 4 49.55 & 0.049 & -30 3 20.4 & 0.046  &  24.0 & 15.9 $^{+2.1}_{-2.1}$ & 3.9 $^{+2.6}_{-1.9}$ & 13.5 $^{+1.7}_{-1.8}$ & 5.8 $^{+1.1}_{-1.0}$ & 8.6 $^{+1.8}_{-1.6}$ & 10.1 $^{+1.3}_{-1.3}$ & L \\
$s02$ & 18 4 48.04 & 0.171 & -30 3 18.9 & 0.114  &  5.1  & 1.5  $^{+0.7}_{-0.6}$ & 3.1 $^{+2.4}_{-1.6}$ & 1.8 $^{+0.7}_{-0.6}$  & 1.4 $^{+0.5}_{-0.4}$ & <0.4 & 1.3 $^{+0.5}_{-0.4}$  & - \\
$s03$ & 18 4 50.69 & 0.070 & -30 3 14.7 & 0.076  &  12.2 & 6.4  $^{+1.4}_{-1.2}$ & <1.2                    & 5.1 $^{+1.1}_{-1.0}$  & 2.7 $^{+0.7}_{-0.6}$ & 2.4 $^{+1.0}_{-0.8}$ & 3.8 $^{+0.9}_{-0.7}$  & L* \\
$s04$ & 18 4 48.76 & 0.126 & -30 3 0.10 & 0.143  &  4.9  & 2.3  $^{+0.9}_{-0.7}$ & <2.3                    & 1.8 $^{+0.7}_{-0.6}$  & 1.1 $^{+0.5}_{-0.4}$ & 0.5 $^{+0.6}_{-0.4}$ & 1.4 $^{+0.5}_{-0.4}$  & -  \\
$s05$ & 18 4 49.99 & 0.101 & -30 3 40.2 & 0.870  &  3.9  & 1.6  $^{+0.8}_{-0.6}$ & 1.1 $^{+1.7}_{-0.8}$ & 1.5 $^{+0.7}_{-0.6}$  & 0.8 $^{+0.4}_{-0.3}$ & 0.8 $^{+0.7}_{-0.5}$ & 1.2 $^{+0.5}_{-0.4}$  & - \\
$s06$ & 18 4 49.36 & 0.131 & -30 3 28.4 & 0.100  &  5.2  & 3.0  $^{+1.0}_{-0.8}$ &   <1.3                 & 2.4 $^{+0.8}_{-0.7}$  & 0.9 $^{+0.5}_{-0.3}$ & 1.7 $^{+0.9}_{-0.7}$ & 1.8 $^{+0.6}_{-0.5}$  & - \\
$s07$ & 18 4 49.29 & 0.105 & -30 3 26.2 & 0.065  &  12.6 & 5.2  $^{+1.3}_{-1.1}$ & 7.4 $^{+3.3}_{-2.6}$ & 5.7 $^{+1.2}_{-1.0}$  & 1.6 $^{+0.6}_{-0.5}$ & 5.2 $^{+1.4}_{-1.2}$ & 4.2 $^{+0.9}_{-0.8}$  & - \\
$s08$ & 18 4 49.82 & 0.120 & -30 3 21.0 &  0.104 &  3.5  & 1.3  $^{+0.8}_{-0.6}$ & 1.7 $^{+2.0}_{-1.2}$ & 1.4 $^{+0.7}_{-0.5}$  & 0.5 $^{+0.4}_{-0.3}$ & 1.1 $^{+0.8}_{-0.5}$ & 1.1 $^{+0.5}_{-0.4}$  & - \\
$s09$ & 18 4 50.90 & 0.106 & -30 3 20.1 & 0.155  &  3.4  & 0.1  $^{+0.4}_{-0.1}$ & 6.2 $^{+2.9}_{-2.2}$ & 1.4 $^{+0.7}_{-0.5}$  & 0.1 $^{+0.2}_{-0.1}$ & 1.8 $^{+0.9}_{-0.7}$ & 1.0 $^{+0.5}_{-0.4}$  & - \\
$s10$ & 18 4 49.37 & 0.107 & -30 3 19.6 &  0.109 &  3.8  & 1.0  $^{+0.7}_{-0.5}$ & 3.9 $^{+2.6}_{-1.9}$ & 1.6 $^{+0.7}_{-0.6}$  & 1.0 $^{+0.5}_{-0.4}$ & 0.4 $^{+0.6}_{-0.4}$ & 1.2 $^{+0.6}_{-0.4}$  & - \\
$s11$ & 18 4 49.62 & 0.133 & -30 3 17.7 &  0.115 &  3.3  & 1.9  $^{+0.9}_{-0.7}$ & <1.3                    & 1.5 $^{+0.7}_{-0.5}$  & 0.9 $^{+0.5}_{-0.3}$ & 0.5 $^{+0.6}_{-0.4}$ & 1.1 $^{+0.5}_{-0.4}$  & - \\
$s12$ & 18 4 49.93 & 0.201 & -30 3 16.7 & 0.104  &  3.6  & 0.6  $^{+0.6}_{-0.4}$ & 4.2 $^{+2.7}_{-1.9}$ & 1.3 $^{+0.6}_{-0.5}$  & 0.7 $^{+0.4}_{-0.3}$ & 0.6 $^{+0.6}_{-0.4}$ & 9.9 $^{+0.5}_{-0.4}$  & - \\
$s13$ & 18 4 48.36 & 0.112 & -30 3 10.4 &  0.071  &  5.3  & 1.7  $^{+0.8}_{-0.6}$ & 4.4 $^{+2.7}_{-1.9}$ & 2.2 $^{+0.8}_{-0.6}$  & 1.4 $^{+0.5}_{-0.4}$ & 0.5 $^{+0.6}_{-0.4}$ & 1.7 $^{+0.6}_{-0.5}$  & - \\
$s14$ & 18 4 45.92 & 0.085 & -30 3 45.7 &  0.078  &  11.6 & 6.0  $^{+1.4}_{-1.2}$ & 6.8 $^{+3.2}_{-2.5}$ & 6.1 $^{+1.2}_{-1.1}$  & 3.9 $^{+0.9}_{-0.8}$ & 1.5 $^{+0.8}_{-0.6}$ & 4.6 $^{+0.9}_{-0.8}$  & - \\
$s15$ & 18 4 47.52 & 0.141 & -30 2 53.9 & 0.127  &  5.5  & 0.2  $^{+0.4}_{-0.2}$ & 7.6 $^{+3.3}_{-2.6}$ & 1.6 $^{+0.7}_{-0.5}$  & 1.1 $^{+0.5}_{-0.4}$ & 0.2 $^{+0.4}_{-0.2}$ & 1.2 $^{+0.5}_{-0.4}$  & - \\
$s16$ & 18 4 48.91 & 0.257 & -30 2 44.0 & 0.084  &  3.4  & 1.3  $^{+0.7}_{-0.5}$ & <1.3                    & 1.0 $^{+0.6}_{-0.4}$  & 0.1 $^{+0.3}_{-0.1}$ & 1.3 $^{+0.8}_{-0.6}$ & 0.8 $^{+0.4}_{-0.3}$  & -\\
$s17$ & 18 4 46.65 & 0.165 & -30 3 56.1 & 0.152 &  3.9  & 1.1  $^{+0.7}_{-0.5}$ & 4.4 $^{+2.7}_{-1.9}$ & 1.8 $^{+0.7}_{-0.6}$  & 1.1 $^{+0.5}_{-0.4}$ & 0.5 $^{+0.6}_{-0.4}$ & 1.3 $^{+0.6}_{-0.4}$  & - \\
$s18$ & 18 4 51.45 & 0.227 & -30 3 40.8 & 0.153  &  5.7  & 1.7  $^{+0.8}_{-0.6}$ & 1.0 $^{+1.7}_{-0.8}$ & 1.6 $^{+0.7}_{-0.6}$  & 0.7 $^{+0.4}_{-0.3}$ & 0.9 $^{+0.7}_{-0.5}$ & 1.2 $^{+0.5}_{-0.4}$  & - \\
[0.1cm]
\hline
\end{tabular}
\begin{tablenotes}
      \centering
      \item {(1) Equatorial coordinates of sources in equinox J2000; 
      (2) the signal-noise ratios; 
      (3) Unabsorbed X-ray flux of 0.5-7~keV in obs1 and obs2; 
      (4) Unabsorbed flux of 0.5-7~keV in the merged observation; 
      (5) Unabsorbed flux of the soft + medium band (0.5–2.0 keV) and the hard band (2.0-7.0 keV) in the merged observation; 
      (3-5): Unit of $10^{-15}$~erg~cm$^{-2}$~s$^{-1}$
      (6) X-ray Luminosity in the merged observation; 
      (7) Flags for indicating the variability of each source: L= long-term variable source, L*= long-term variability candidate with an upper limit in one epoch, but a flux difference that still exceeds the variability threshold.}
    \end{tablenotes}
\end{table*}

\subsection{Spectral analysis}
For the four brightest X-ray sources from the 18 sources (s01, s03, s07, and s14), we further investigated their emission nature by examining their X-ray spectra. We utilized the \texttt{CIAO} tool \texttt{specextract} to extract the spectra and compute the response files. This tool creates source and background spectra along with response matrix files (RMFs) and ancillary response files (ARFs). For sources with detections in both observations, spectra were extracted separately for each epoch. We used the task \texttt{combine\_spectra} to sum the spectral counts and exposure times from the two observations, creating a single spectrum for each source with improved S/N. 

Spectral fitting was performed with Xspec (version 12.14.0b). Two emission models were tested: a power-law model tbabs(\textit{powerlaw}), typically associated with non-thermal emission processes, and a blackbody model tbabs(\textit{bbody}), indicative of thermal emission. We fit models with both a fixed $N_H$ (3.7 $\times$ 10 $^{21}$ cm$^{-2}$). The best-fit parameters for each source are presented in \autoref{tab:spec_fits}. Errors are quoted at the 1$\sigma$ confidence level for a single parameter of interest.

For the fixed $N_H$ fits, there was no significant difference in the quality of the model fits between the two models, except for s14, where the power-law model was preferred ($\chi^2/\nu = 8.6/11$ vs 14.0/11). The best-fit photon indices for the power-law fits are consistent with the X-ray colours plotted in \autoref{fig:xray_CMD}, with s07 showing a hard spectrum ($\Gamma = 0.8\pm0.4$). 

We also tested spectral fits allowing $N_H$ to vary, but these fits resulted in larger uncertainties and poorly constrained parameters. Overall, because of the high foreground $N_H$ and faintness of the X-ray sources, these spectral fits provide only limited constraints on their intrinsic physical properties.

\begin{table*}
\centering
\caption{Results of \textit{Chandra} X-ray spectral fits for the brightest sources in NGC 6528 using power-law and blackbody models. The fits were performed under two scenarios: fixed absorption and free absorption.}
\label{tab:spec_fits}
\renewcommand{\arraystretch}{1.3}
\begin{tabular}{c|cccc|cccc}
\toprule
 & \multicolumn{4}{c|}{\textbf{Power-law}} 
 & \multicolumn{4}{c}{\textbf{BlackBody}} \\
\midrule
{Source} & $N_H$ & $\Gamma$ & L$_x$ & $\chi^2/\nu$ & $N_H$ & kT & L$_x$ & $\chi^2/\nu$ \\
& & & \scriptsize {($10^{31}$~erg~s$^{-1}$)} & & & \scriptsize{(keV)} & \scriptsize {($10^{31}$~erg~s$^{-1}$)} &  \\
\midrule
\multicolumn{9}{c}{fixed $N_H$ \scriptsize (0.37 $\times$ 10 $^{22}$ cm$^{-2}$)} \\
\midrule
s01 & Fixed & $1.6 \pm 0.3$ & $11.1$$ ^{+0.2}_{-0.2}$ & $15.0/12$ & Fixed & $0.7 \pm 0.1$ & $9.0$$ ^{+1.4}_{-1.4}$ & $13.8/12$ \\
s03 & Fixed & $2.5 \pm 0.5$ & $3.2$$ ^{+0.8}_{-0.7}$ & $14.0/10$ & Fixed & $0.5 \pm 0.1$ & $2.8$$ ^{+0.5}_{-1.1}$ & $14.8/10$ \\
s07 & Fixed & $0.8 \pm 0.4$ & $6.9$$ ^{+1.3}_{-2.1}$ & $4.6/12$  & Fixed & $1.0 \pm 0.2$ & $5.9$$ ^{+1.0}_{-1.5}$ & $3.1/12$   \\
s14 & Fixed & $2.8 \pm 0.5$ & $3.0$$ ^{+0.8}_{-0.8}$ & $8.6/11$  & Fixed & $0.36 \pm 0.04$ & $2.2$$^{+0.5}_{-0.6}$ & $14.0/11$ \\
\midrule
\multicolumn{9}{c}{free $N_H$ \scriptsize(10$^{22}$ cm$^{-2}$) } \\
\midrule
s01 & $0.94 \pm 0.47$ & $2.3 \pm 0.7$ & $10.0$$ ^{+1.3}_{-5.2}$ & $13.7/11$& $0.08 \pm 0.32$ & $0.8 \pm 0.1$  & $9.4$$ ^{+1.1}_{-1.9}$ & $13.1/11$ \\
s03 & $0.76 \pm 0.54$ & $3.2 \pm 0.1$ & $2.8$$ ^{+0.1}_{-0.0}$ & $13.2/9$ & $0.05 \pm 0.37$ & $0.6 \pm 0.1$  & $2.8$$ ^{+0.2}_{-1.3}$ & $13.9/9$  \\
s07 & $1.33 \pm 0.97$ & $1.7 \pm 1.0$ & $6.2$$ ^{+0.2}_{-0.0}$ & $3.2/11$ & $0.41 \pm 0.65$ & $1.0 \pm 0.3$  & $5.9$$ ^{+0.9}_{-2.0}$ & $3.1/11$  \\
s14 & $0.15 \pm 0.37$ & $2.1 \pm 1.2$ & $3.7$ $ ^{+0.6}_{-0.0}$ & $8.2/10$ & - & - & - & - \\
\bottomrule
\end{tabular}
\end{table*}

\section{Optical Counterpart Analysis} 
\subsection{Hubble Space Telescope} 
We used archival images made with the \textit{HST} Wide-Field Camera 3 (WFC3) and the Advanced Camera for Surveys (ACS) covering the half-light radius of NGC 6528. We used data in F390W(U$_{390}$) (2206 s), F555W(V$_{555}$) (1381 s), and F814W(I$_{814}$) (791s) for WFC3, and F606W(V$_{606}$) (504 s) and F814W(I$_{814}$) (371 s) for ACS. The WFC3 observations were done in 2010, approximately 8 years after the ACS. To obtain the photometry data of each source present, we used the Hubble Source Catalog (HSC) V3 (\citealt{Whitmore_2016}). The details of \textit{HST} data used in this study are summarized in \autoref{tab:HSTdata}.

\begin{table}
	\centering
         \begin{tabular}{ccccc}
 \hline
Prop ID & Obs start date & Exp. & Inst. & Filter \\
&&\scriptsize(s)&&
\\
\hline
11664 & 2010-06-26 & 2206 & WFC3/UVIS & F390W \\
11664 & 2010-06-26 & 1381 & WFC3/UVIS & F555W \\
9453  & 2002-08-06 & 504  & ACS/WFC   & F606W \\
11664 & 2010-06-26 & 791  & WFC3/UVIS & F814W \\
9453  & 2002-08-06 & 371  & ACS/WFC   & F814W \\
\hline
\end{tabular}
    \caption{\textit{HST} data used in this study}
        \label{tab:HSTdata}
\end{table}

\subsubsection{Methodology}
\begin{figure*}
    \centering
    \includegraphics[width=0.99\textwidth]{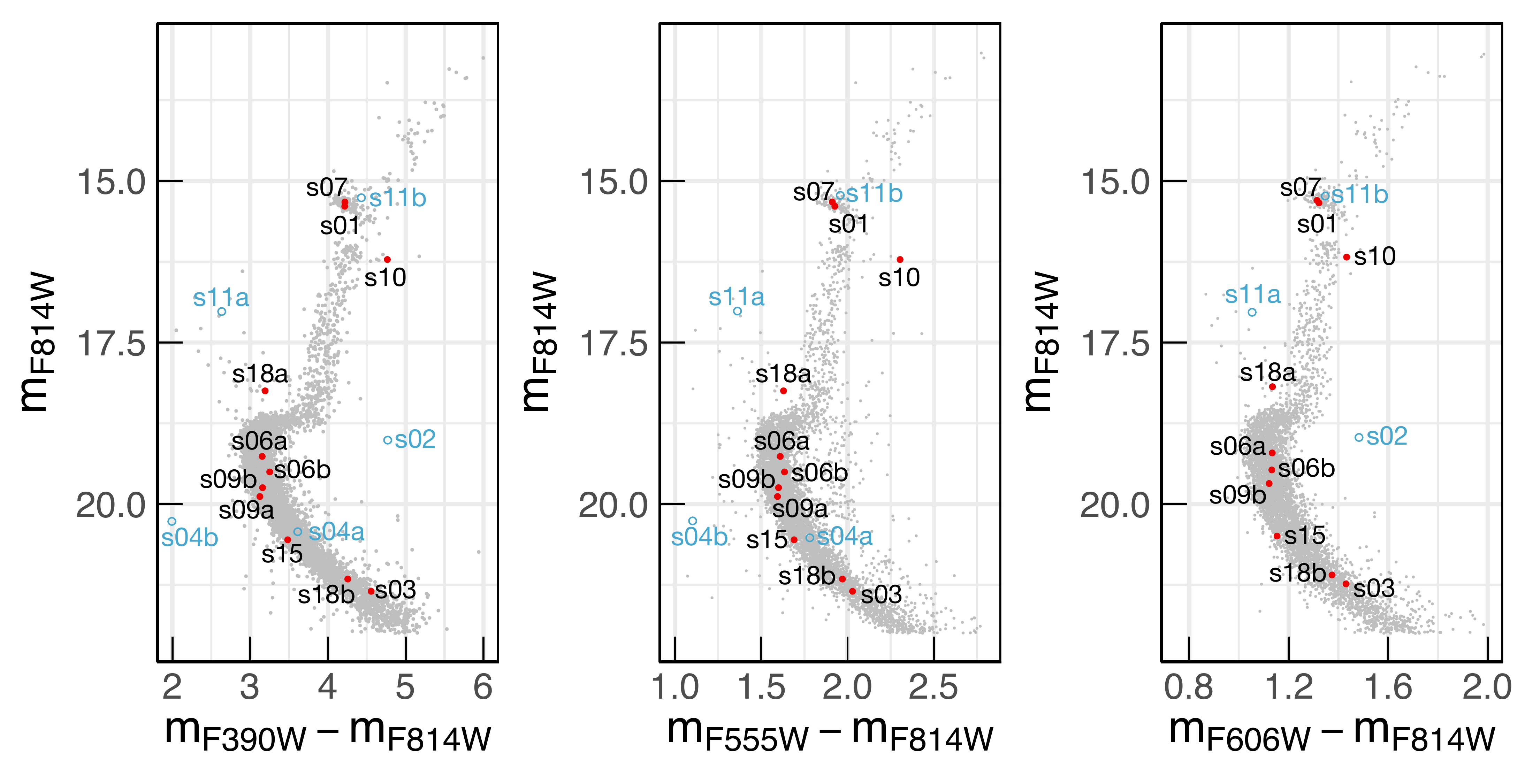}
    \caption{Reddening-corrected CMDs of NGC 6528 constructed using \textit{HST} photometry for 5,867 stars located within the cluster's half-light radius. Symbols mark sources that fall within the 1$\sigma$ positional uncertainties of the \textit{Chandra} X-ray detections. Red dots indicate cluster-member counterparts to X-ray sources, while blue open circles represent additional candidates that are not cluster members but spatially coincide with X-ray positions.}
    \label{fig:CMD_combined}
\end{figure*}

We initially compiled a photometric catalogue containing 19,490 sources. To divide these sources into cluster members and field stars, we cross-matched our photometric catalogue with the kinematic membership list of \citet{Lagioia_2014}. From this list, we identified a total of 5,867 cluster members located within the half-light radius of NGC 6528. To identify optical counterparts, we required that sources lie within the X-ray's 1$\sigma$ error region as determined by \texttt{wavdetect}. This was chosen to minimize confusion from source crowding since bigger error regions (2$\sigma$ and 3$\sigma$) often had more than 10 sources within them. 
For comparison, we also computed 95\% confidence positional uncertainties following of \citet{Kim_2007}. Across our sample, we find that the \texttt{wavdetect} 1$\sigma$ regions are on average $\sim$27\% smaller than the \citet{Kim_2007} values. Given this comparison, we adopt the \texttt{wavdetect} 1$\sigma$ error ellipses throughout our analysis.
For these sources, we constructed reddening-corrected CMDs (see \autoref{fig:CMD_combined}) to facilitate the analysis of potential optical counterparts. In cases where multiple optical candidates were associated with a single X-ray source region, we appended distinguishing letters to their IDs to avoid ambiguity. After these stages, we found a total of 11 HST matching sources within the X-ray 1$\sigma$ error regions. The final list of sources and their photometric properties is provided in \autoref{tab:hst_data}.

\begin{table*}
\centering
\renewcommand{\arraystretch}{1.3}
\begin{tabular}{cc|ccc|cc|c|c|c|c}
\hline
\multicolumn{2}{c|}{ID} & 
\multicolumn{3}{c|}{WFC3 Photometry} &
\multicolumn{2}{c|}{ACS Photometry} &
\multicolumn{1}{c|}{C$_{dist}$*} &
\multicolumn{1}{c|}{P} &
\multicolumn{1}{c|}{N} & 
\multicolumn{1}{c}{B} \\
{Source} & {$HST$ ID} & $U_{390}$ & $V_{555}$ & $I_{814}$ & $V_{606}$ & $I_{814}$ & {(arcsec)} & & \\
\hline
s01  & 73336848  & 19.61 & 17.32 & 15.39 & 16.66 & 15.34 & 0.32 & 0.36 & 1 & 1.82 \\
s03  & 35737640  & 25.90 & 23.38 & 21.35 & 22.67 & 21.23 & 0.19 & 0.47 & 1 & 1.15\\
s06a & 3471366   & 22.42 & 20.87 & 19.26 & 20.34 & 19.21 & 0.28 & 0.25 & 2 & 1.96\\
s06b & 91105334  & 22.75 & 21.14 & 19.50 & 20.60 & 19.47 & 0.10 & 0.25 & 2 & 1.96\\
s07  & 19294730  & 19.54 & 17.24 & 15.33 & 16.62 & 15.30 & 0.44 & 0.26 & 1 & 2.91\\
s09a & 14911830  & 23.01 & 21.48 & 19.88 & 20.92 & 31.76 & 0.17 & 0.34 & 2 & 0.99\\
s09b & 88388114  & 22.91 & 21.35 & 19.75 & 20.80 & 19.68 & 0.22 & 0.34 & 2 & 0.99\\
s10  & 27165631  & 20.98 & 18.52 & 16.22 & 17.61 & 16.18 & 0.10 & 0.44 & 1 & 1.27\\
s15  & 58583478  & 24.04 & 22.24 & 20.55 & 21.65 & 20.49 & 0.26 & 0.51 & 1 & 0.98\\
s18a & 12731840  & 21.44 & 19.88 & 18.25 & 19.32 & 18.18 & 0.36 & 0.24 & 2 & 2.13\\
s18b & 103848938 & 25.41 & 23.13 & 21.16 & 22.47 & 21.10 & 0.69 & 0.24 & 2 & 2.13\\
\hline
\end{tabular}
\begin{tablenotes} 
      \centering
      \item *Cdist is the distance from the optical source to the centre of the X-ray region.
\end{tablenotes}
\caption{Summary of \textit{HST} Optical Counterparts}
\label{tab:hst_data}
\end{table*}

\subsubsection{Probability analysis}
We have also assessed the probability of a counterpart being a true counterpart or a chance of positional coincidence. First, we took the original 1$\sigma$ elliptical error regions of \textit{Chandra} and converted them to circles of equivalent area for consistency. Subsequently, we generated 10,000 circular regions uniformly distributed within an annular region with an inner radius of 1.5" and an outer radius of 4.0" from the centre of the original X-ray detection. The inner boundary was selected to ensure no overlap between the simulated circles and the X-ray regions, while the outer boundary limits the analysis to around the vicinity of the X-ray source. 

For each simulated circle, we counted the number of $HST$ sources within its outer region boundary. We then computed the average background number (B) of $HST$ sources by dividing the total number of sources counted by the number of simulations (10,000). The probability that any of the $HST$ source within an X-ray region is a true counterpart was calculated as:
\[
P = \frac{N}{N+B}
\]
where B is the average number of background stars and N is the number of observed $HST$ sources inside the X-ray regions. We treated N as the sum of a Bernoulli-distributed counterpart and a Poisson-distributed background source population, assuming a prior probability of 0.5 of a true counterpart source being present in the data. This probability represents whether any of the detected $HST$ sources is a counterpart; if $N > 1$, then the probability of an individual source being the counterpart is $1/(N+B)$. The results from this probability analysis are summarized in \autoref{tab:hst_data}. 

\begin{figure*} 
    \centering
    \includegraphics[width=\textwidth]{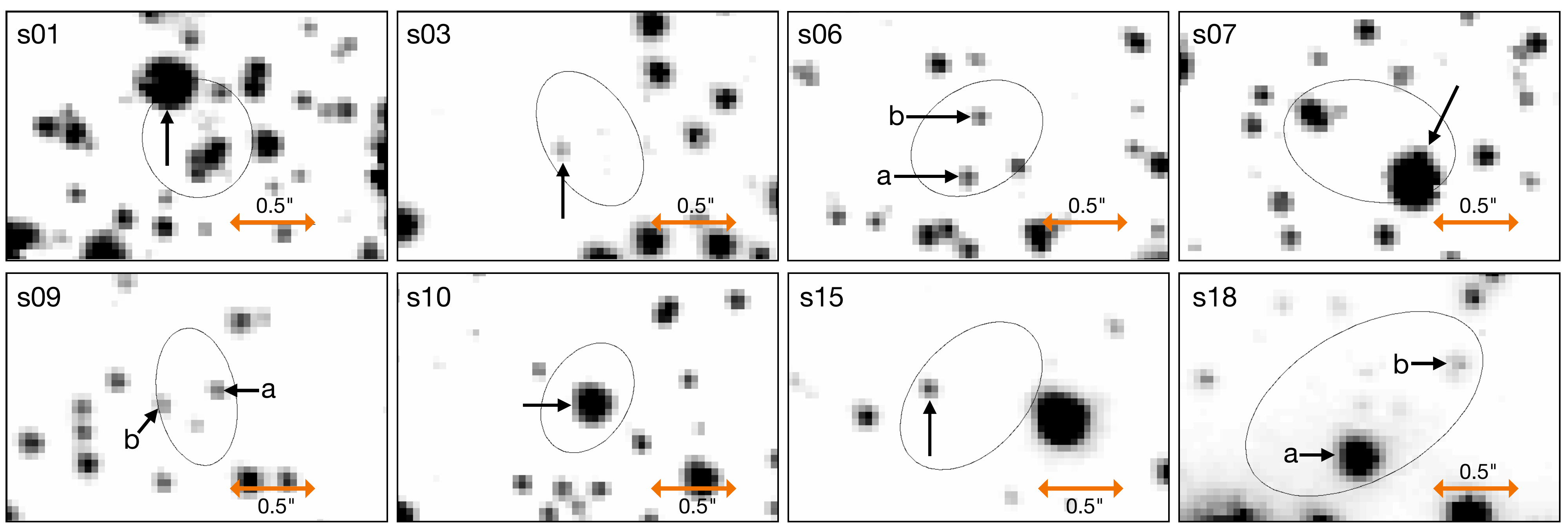}
    \caption{\textit{HST} Finding chart of candidate optical counterparts within the half-light radius of NGC 6528. X-ray source positions are marked with 1$\sigma$ error ellipses, and corresponding candidate optical counterparts are indicated by arrows. Each panel is labeled with the source ID, and the scale bars denote 0.5 arcseconds.}
    \label{fig:hst_images}
\end{figure*}

\subsection{\textit{Gaia}}
\subsubsection{Gaia counterpart with \textit{Chandra} X-ray sources}
\begin{figure}
    \centering
    \includegraphics[width=0.9\columnwidth]{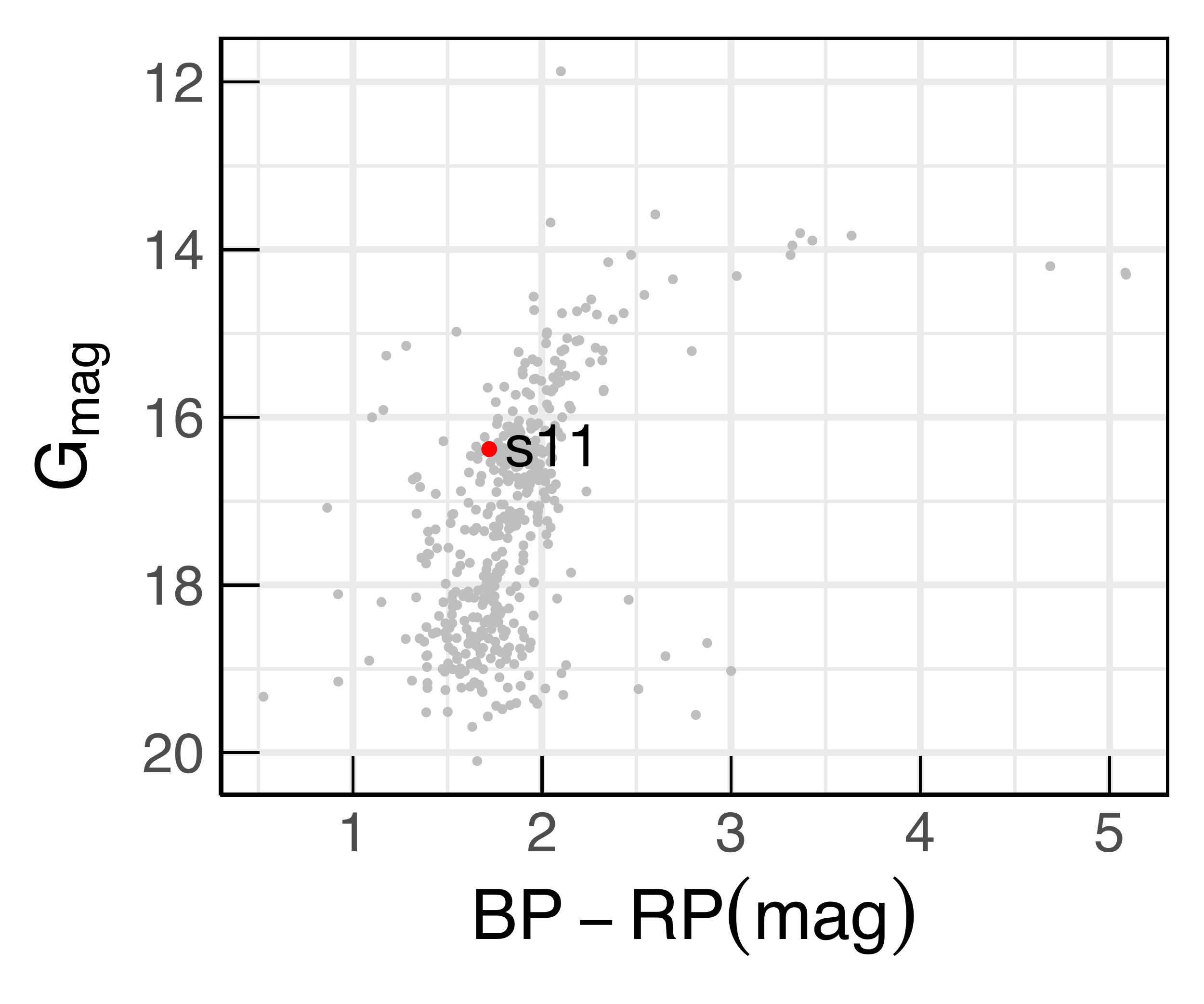}
    \caption{CMD of \textit{Gaia} DR3 sources within the half-light radius, showing the G-band magnitude as a function of BP-RP colour. A total of 477 sources with valid photometry are included.}
    \label{fig:gaia_cmd}
\end{figure}

In addition to the \textit{HST} analysis above, we also searched for potential optical counterparts in the \textit{Gaia} DR3 catalog \citep{GAIA_2023} to find any specific evidence from the \textit{Gaia} proper motion measurements. Within the half-light radius of NGC 6528, we identified a total of 1,259 \textit{Gaia} sources. We then searched for matches to X-ray sources by examining the 1$\sigma$ elliptical positional uncertainty regions around each \textit{Chandra} detection. While five \textit{Gaia} sources (s07, s08, s10, s11, and s14) fall within the \textit{Chandra} error ellipses, only those associated with s07 and s08 have proper motion information available in DR3. We also calculated the probability of association for each \textit{Gaia} candidate using the same method described in Section 4.1.2. Since \textit{Gaia} sources are less densely distributed than those from \textit{HST}, the resulting probabilities are generally higher, indicating a greater likelihood that these sources are true counterparts. \autoref{fig:gaia_cmd} presents the CMD of \textit{Gaia} DR3 sources located within the half-light radius of NGC 6528. A total of 477 sources have valid photometric measurements in both the BP and RP bands and are included in the CMD. Among our X-ray counterparts, only s11 has BP/RP colour information and is located near the main-sequence track.

\subsubsection{\textit{Gaia} Proper Motion Analysis}
Additionally, we analysed the proper motion of each source from \textit{Gaia} DR3. From 1,259 \textit{Gaia} detected sources, 640 sources have information on their proper motion and their associated errors. We employed an extreme deconvolution (XD) technique \citep{Bovy_2011} integrated within an expectation-maximization (EM) framework \citep{EM_algorithm_1977} to robustly deconvolve measurement uncertainties in the \textit{Gaia} DR3 data to assign probabilistic cluster membership. This approach treats measurement uncertainties directly in the likelihood and iteratively refines both cluster and field parameters.

In our two-component Gaussian mixture model, each star \(i\) has a measured proper motion \(\mu_i\) with a known uncertainty covariance \(\boldsymbol{S}_i\). Let \(\boldsymbol{\mu}_\mathrm{c}\) and \(\boldsymbol{\mu}_\mathrm{f}\) denote the mean proper motions of the cluster and field components, with covariance matrices \(\boldsymbol{\Sigma}_\mathrm{c}\) and \(\boldsymbol{\Sigma}_\mathrm{f}\), and mixing fractions \(\pi_\mathrm{c}\) and \(\pi_\mathrm{f}\). The likelihoods for the cluster and field components are defined as
\[
L_c =\mathcal{N}\Bigl(\mu_i \mid \mu_c,\, \Sigma_c + S_i\Bigr)
\]
\[
L_f =\mathcal{N}\Bigl(\mu_i \mid \mu_f,\, \Sigma_f + S_i\Bigr)
\]
and the membership probability is given by
\[
P(\mathrm{member} \mid \mu_i) = \frac{\pi_c\,L_c}{\pi_c\,L_c + \pi_f\,L_f}.
\]

We then apply the EM algorithm to iteratively update the model parameters until convergence. After convergence, the stars with a membership probability exceeding a chosen threshold (e.g., $P \geq 0.7$) are classified as likely cluster members.

\begin{figure}
    \centering
    \includegraphics[width=0.9\columnwidth]{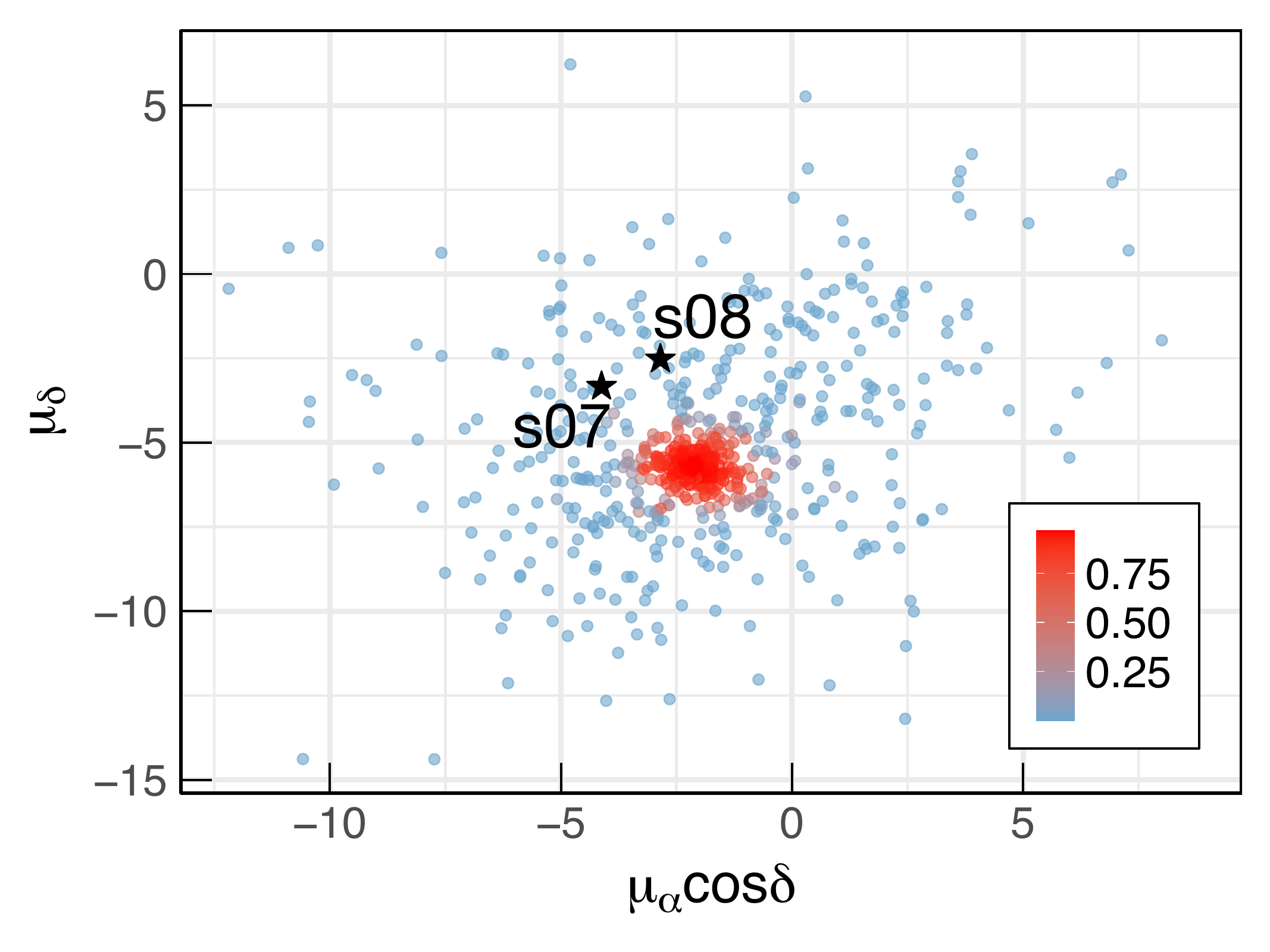}
    \caption{Proper-motion distribution of \textit{Gaia} DR3 sources within the half-light radius of NGC 6528, plotted in the 2-dimensional proper motion plane. Each point is colour-coded by its cluster membership probability, as derived from our extreme deconvolution analysis.}
    \label{fig:2dgaussian}
\end{figure}

In \autoref{fig:2dgaussian}, we present the proper motion distribution of \textit{Gaia} DR3 sources, with each star colour-coded according to its cluster membership probability. The redder colour represents a higher likelihood of cluster membership. The XD-EM algorithm effectively deconvolves measurement uncertainties, revealing a well-defined cluster population. Notably, both s07 and s08 lie outside the range of the cluster distribution and are thus unlikely to be cluster members based on their kinematics.

\begin{table*}
	\centering
        \renewcommand{\arraystretch}{1.3}
         \begin{tabular}{ccccccccccccc}
 \hline
{ID$^{(1)}$} & {Gaia ID$^{(2)}$} & $\mu_{\alpha}\cos\delta^{(3)}$ & $\sigma_{\alpha}^{(4)}$ & $\mu_{\delta}^{(3)}$ & $\sigma_{\delta}^{(4)}$ & $\rho_{\mu_{\alpha},\mu_{\delta}}^{(5)}$ & ${G_{mag}}^{(6)}$ & $R^{(7)}$
& P$^{(8)}$ \\
& (\textit{Gaia} DR3) & $(mas~yr^{-1})$ & &  $(mas~yr^{-1})$ & & & & {(arcsec)}  &
\\
\hline
s07 & 4050189348856775808 & -4.12 & 0.17 & -3.33 & 0.12 & 0.12 & 16.48 & 0.41 & 0.85 \\
s08 & 4050189348856776320 & -2.85 & 0.28 & -2.52 & 0.20 & 0.18 & 16.50 & 0.23 & 0.91 \\
s10 & 4050189348856759936 & -     & -    & -     &   -  & -    & 17.09 & 0.11 & 0.93 \\
s11 & 4050189348856756992 & -     & -    & -     &   -  & -    & 16.38 & 0.20 & 0.95 \\
s14 & 4050188970899724672 & -     & -    & -     &   -  & -    & 17.16 & 0.33 & 0.94 \\
\hline
\end{tabular}
\begin{tablenotes} 
      \centering 
      \item {(1) Corresponding \textit{Chandra} detected source ID; (2) Gaia DR3 designation ID (3) proper motion vector (4) proper motion errors for $\mu_{\alpha}\cos\delta$ and $\mu_{\delta}$ \\(5) Correlation coefficient between $\mu_{\alpha}\cos\delta$ and $\mu_{\delta}$ (6) G-Band Mean Magnitude measured by \textit{Gaia} \\(7) Angular distance between \textit{Chandra} and \textit{Gaia}. (8) Probability of Gaia source within \textit{Chandra} region is a true counterpart.}
    \end{tablenotes}
    \caption{Proper motions and photometry of \textit{Gaia} counterparts to \textit{Chandra} sources.}
        \label{tab:GAIA}
\end{table*}

\subsubsection{\textit{Gaia} and \textit{HST} within 1$\sigma$ \textit{Chandra} X-ray detection}
To further identify the optical counterparts associated with the X-ray source, we conducted a positional cross-match between the \textit{HST} cluster member list and the \textit{Gaia} DR3 catalog. We initially selected possible cluster sources within the half-light radius of 0.91 arcmin. To ensure a robust association, we then adopted a matching criterion of 0.1 arcsec between \textit{HST} cluster member list and \textit{Gaia} sources.

Without requiring \textit{Gaia} proper-motion measurements ($\mu_{\alpha}cos\delta$, $\mu_{\delta}$), we found 465 candidate matches within 0.1 arcsec of the \textit{HST} positions. Of these, two sources matched our 1$\sigma$ X-ray detection region (s07 and s10).
When we restricted the \textit{Gaia} sample only to sources with valid proper motions, we obtained 289 matches between \textit{HST} and \textit{Gaia}. Among these, only s07 coincided with our \textit{Chandra} X-ray position that lies on the red giant branch (RGB). Notably, s07 appears in all of our available CMDs (see \autoref{fig:CMD_combined}). We note that while s07 is classified as a cluster member in the \textit{HST} cluster member list by \citet{Lagioia_2014}, our \textit{Gaia}-based proper motion analysis indicates it is a non-member. However, the renormalized unit weight error (RUWE) for the Gaia astrometric fit is poor (RUWE=2.2), potentially indicating that the proper motion is unreliable; its membership can be revisited after a future Gaia data release.

\subsection{Supplementary data by other catalogs}
Based on observations from the \textit{La Silla} Observatory, we found two positional coincidences were identified for our X-ray sources s06 and s10. They were classified by \citet{Skottfelt_2015} as an eclipsing binary and an RR Lyrae star (RR0, i.e. fundamental-mode), respectively. In our notation, s06 corresponds to V7 (in \citealt{Skottfelt_2015}), and s10 corresponds to V2, and s10 has a period of 0.8165 days, which is typical for a fundamental-mode RR Lyrae pulsator.

Additionally, we identified two more X-ray sources (s11 and s14) with counterparts in the \textit{OGLE-IV} database \citep{Udalski_2015}. Specifically, s11 corresponds to OGLE-BLG-ECL-290997, and s14 corresponds to OGLE-BLG-ECL-290300. Both are classified in the \textit{OGLE} catalog as eclipsing binaries, with reported periods of 1.029 days (s11) and 1.048 days (s14). We confirmed these associations by matching the \textit{OGLE} catalog coordinates to the \textit{Chandra} detections within the 1$\sigma$ positional uncertainty regions. 

\subsection{X-ray/optical flux ratio}
To investigate the nature of X-ray sources with identified optical counterparts, we compare each source's X-ray luminosity (0.5--2.5 keV) with its optical absolute magnitude ($M_V$). For consistency, we recalculate the X-ray luminosity within the range of 0.5--2.5 keV using \texttt{srcflux} described in Section~2.2. For sources with multiple potential counterparts, we consider each in turn. We use $V_{555}$ band photometry and apply both reddening and extinction corrections to derive $M_V$, adopting the cluster distance from \citet{Harris_2010} and $R_V$ from \citealt{Schlafly_2011}. We then employ the empirical relations from \citet{Bassa_2004} and \citet{Verbunt_2008} to guide the source classification. 

\begin{figure}
    \centering
    \includegraphics[width=\columnwidth]{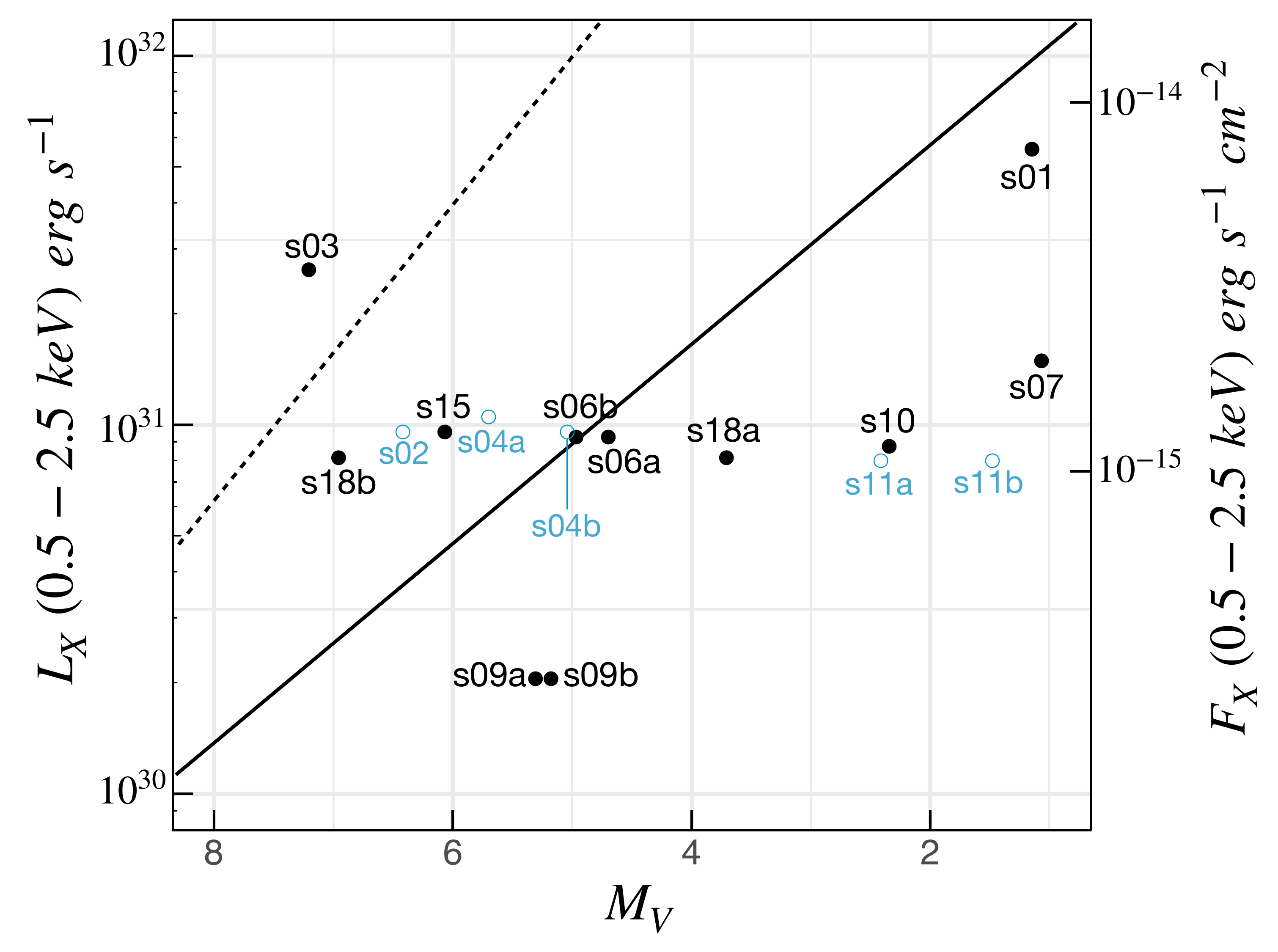}
    \caption{X-ray to optical flux ratio diagram for \textit{Chandra} sources in NGC 6528 with identified \textit{HST} counterparts. Black dots mark cluster members; open blue circles indicate non-members within the X-ray positional uncertainty. X-ray luminosities are in the 0.5–2.5~keV band, and optical magnitudes are $M_V$ from $V_{555}$.}
    \label{fig:ratio}
\end{figure}

In \autoref{fig:ratio}, the dashed line ($ logL_X=34.0-0.4M_V$) represents an approximate boundary between CVs and ABs \citep{Bassa_2004}, while the solid line ($logL_X=32.3-0.27M_V$) sets an upper limit of coronal emission for ABs \citep{Verbunt_2008}. In general, sources lying well above the dashed line are tend to be CVs that have $L_X\sim10^{31-33}$~erg~s$^{-1}$, and those below the solid line are consistent with ABs. However, the sources lying in the intermediate region can be either faint CVs, bright ABs, or MSPs. Furthermore, background active galactic nuclei can also appear with CV-like characteristics if incorrectly matched to a cluster optical source. 

In this study, we utilize the flux-ratio diagram together with our earlier multi-wavelength analyses to enhance the accuracy of source classification, facilitating deeper insight into the nature of exotic binary systems within this cluster. We note that these flux ratios are most informative when the X-ray and optical identifications are firmly established.

\section{Discussion}
\subsection{Individual source analysis}
\subsubsection{s01}
s01 is the brightest X-ray source we detect within the half-light radius of NGC 6528, with an unabsorbed X-ray luminosity $L_X\sim10^{32}$~erg~s$^{-1}$. It also exhibits long-term variability between two epochs of the \textit{Chandra} observations. Such variability could reflect fluctuating accretion rates (as often seen in CVs or quiescent low-mass X-ray binaries (qLMXB)) or flare-like outbursts in an unusually luminous AB. However, there is no short-term variability within the single observation length. Its X-ray spectrum can be fitted with either a power-law or a blackbody.

We identify a single \textit{HST} source within the $1\sigma$ positional uncertainty of the X-ray detection, and this source is confirmed as a cluster member based on its location on the RGB across all CMDs (U${390}$, V${555}$, I$_{814}$; see \autoref{fig:CMD_combined}). If this object is the counterpart to the X-ray source, it is plausibly an RS CVn-type AB near the maximum X-ray luminosity expected for such systems.

Although multiple stars are visible in the X-ray error region in \autoref{fig:hst_images}, only one cluster member from \cite{Lagioia_2014} is detected in the \textit{HST} data. The other visually apparent stars are either classified as non-members or are too faint to have reliable proper motion measurements; while we cannot fully exclude the latter as potential counterparts, their membership status remains unconstrained, making the identified cluster member the most plausible counterpart at this time. The long-term X-ray variability and central location of s01 are consistent with expectations for a mass-segregated compact binary. In addition, since NGC 6528 has been associated with Fermi $\gamma$-ray emission \citep{deMenezes2023}, the presence of MSPs is plausible. Follow-up time-series optical photometry or deeper X-ray observations would be useful to better constrain the nature of this source.

\subsubsection{s03} 
s03 is one of the brighter X-ray sources in our sample, with $L_X \sim 4 \times 10^{31}$~erg~s$^{-1}$. It was detected in the first \textit{Chandra} observation but not in the second, implying a significant (factor of 5 or more) flux drop. 

The spectral fits of s03 are statistically acceptable with both power-law and blackbody models. A single \emph{HST} source falls within the s03 error circle, with detections in V$_{555}$, V$_{606}$, and I$_{814}$. From the CMDs (see \autoref{fig:CMD_combined}), this counterpart appears at the bottom of the main sequence region. In \autoref{fig:ratio}, s03 clearly resides above the region typically occupied by ABs (this is even more true if the optical counterpart is fainter than in the HST dataset). Hence, a CV is the most likely classification for this course, though given the X-ray spectrum, a qLMXB cannot be ruled out.

\subsubsection{s06}
s06, with $L_X \sim 2 \times 10^{31}$ erg s$^{-1}$ is a long-term variable candidate only detected in the first \textit{Chandra} observation. We find two possible \textit{HST} optical counterparts (s06a and s06b), and both lie on the main-sequence region of the CMDs (\autoref{fig:CMD_combined}). From ground-based optical observations at the \textit{La Silla} Observatory, \cite{Skottfelt_2015} identified a variable source (their V7) in this region, classifying it as a possible eclipsing binary. With the available data, it is not possible to associate V7 conclusively with either s06a or s06b. In \autoref{fig:ratio}, both s06a and s06b are around the line representing the upper limit for coronal AB emission, so the source is plausibly an AB, but a CV cannot be definitively ruled out without more optical data.

\subsubsection{s07}
s07 is the third brightest X-ray source in our sample, with $L_X \sim 4 \times 10^{31}$ erg s$^{-1}$, and has no detected X-ray variability. Its X-ray spectrum is well-fit by either a hard power-law ($\Gamma=0.8\pm0.4$) or a $\sim1$ keV blackbody. Two \textit{HST} sources (s07a and s07b) lie within the 1 $\sigma$ X-ray positional uncertainty. Both are detected in U$_{390}$, V$_{555}$, V$_{606}$, and I$_{814}$. In the \textit{HST} CMD, s07a appears on the RGB, whereas s07b appears just above the main-sequence turnoff, as a likely subgiant. In \autoref{fig:ratio}, both counterparts lie in the regime usually associated with ABs. However, such a hard X-ray spectrum would be unusual for an AB, especially this X-ray luminosity. Hence, it is possible that neither s07a or s07b is the counterpart and instead it is a fainter star, and the source could be a CV. 

\subsubsection{s10}
s10 is detected in both epochs of \textit{Chandra} and does not show significant variability, with $L_X \sim 10^{31}$ erg s$^{-1}$. In the optical, there is a single \textit{HST} candidate falling within the 1$\sigma$ error of \textit{Chandra} X-ray detection region. Previous ground-based observations \citep{Skottfelt_2015} found evidence for variability and gave a preliminary classification as an RR Lyrae variable star, though they state the variability amplitude is not fully consistent with this classification. However, the \textit{HST} CMD places s10 on the lower RGB rather than in the instability strip where RR Lyraes would be located, suggesting it is not an RR Lyrae. 
Furthermore, in the X-ray/optical flux ratio diagram (\autoref{fig:ratio}), s10 occupies the region normally associated with ABs. Hence, it is likely that this source is an RS CVn star with associated variability due to starspots and/or flares. Time-series optical data would help verify this tentative classification.

\subsubsection{s14} s14 is the second brightest X-ray source within the half-light radius of NGC 6528 ($L_X = 4.6\times10^{31}$~erg~s$^{-1}$). It was detected in both \textit{Chandra} observations, with no significant short or long-term X-ray variability. Spectral analysis reveals that the X-ray emission of this source is relatively soft, with $\Gamma = 2.8\pm0.5$ for the fixed $N_H$ power-law fit; this was also the only source for which a power-law was clearly preferred over a thermal model. We additionally applied an absorbed diffuse plasma model (APEC). Assuming a fixed $N_H$ = 3.7 $\times$ 10 $^{21}$ cm$^{-2}$, the best-fit temperature was $kT = 2.15 \pm 0.64$ keV with the $\chi^2$ value of 12.67 with 11 degrees of freedom. This fit suggests that the X-ray emission from s14 may plausibly originate from a hot, low-density plasma component. A single candidate optical counterpart detected by \textit{HST} V3 catalog lies within the 1$\sigma$ positional uncertainty region from \textit{Chandra}, however, the cluster membership list from \cite{Lagioia_2014} does not include this source. This star is also identified as an eclipsing binary (OGLE-BLG-ECL-290300) in the \textit{OGLE-IV} catalog \citep{Soszynski2016}, with a period of 1.048 d. While this star is also present in Gaia DR3, there is no proper motion listed, making it impossible to confirm its cluster membership in this manner. Given the location of the source near the half-light radius of the cluster, it is possible or even likely that this is a field star (which, if a close eclipsing binary, might indeed be associated with the X-ray source). A proper motion measurement from a future Gaia catalog could determine its cluster membership.

\subsection{Other sources}
The majority of the X-ray sources not discussed individually above are likely to be members of NGC 6528. To estimate the number of background sources within the cluster region, we analyzed an annulus between 2 and 3.5 arcmin from the cluster center, detecting 35 sources (S/N > 3) over 25.9 arcmin$^2$. This yields a background density of $\sim 1.35$ sources arcmin$^2$, implying $\sim 3.5$ expected background sources within the half-light radius. Since 18 sources are detected in this region, we estimate that $\sim 80\%$ of the sources are likely genuine cluster members. However, their relatively low X-ray luminosities, the weak constraints on their X-ray spectra, and the uncertainty in their optical associations all make it difficult to classify these sources, other than the broad statement that they are probably mostly CVs and/or ABs.

Additionally, crowding and detection limits in the \textit{HST} images may prevent the recovery of fainter or blended counterparts, further complicating identification. In the absence of distinctive CMD positions, short-term variability, or characteristic X-ray spectra, a range of compact binary scenarios—including quiescent neutron stars, accreting white dwarfs, or MSP systems—cannot be ruled out. However, such interpretations require caution, as these scenarios often rely on indirect indicators.

Therefore, while most sources are broadly consistent with ABs or CVs, higher signal-to-noise X-ray observations, deeper optical photometry, and time-domain studies will be essential to securely classify the unresolved population and to assess whether more exotic systems may be hidden among them.

\subsection{Supplementary counterpart candidates with uncertain membership}
In addition to the primary sample of X-ray sources with likely cluster member counterparts, we identified several additional candidate counterparts located within the 1$\sigma$ positional uncertainties of \textit{Chandra} detections that are not classified as cluster members by \cite{Lagioia_2014}. These sources, marked as blue open circles in \autoref{fig:CMD_combined} and \autoref{fig:ratio}, are likely field stars based on kinematic and photometric criteria, but their photometric and X-ray properties warrant brief discussion as potential counterparts.

\subsubsection{s02}
s02 has a single HST counterpart located on the red side of the main sequence, significantly below the subgiant branch in the CMD. In the X-ray/optical flux ratio plot, it falls within a region often associated with CV. If a cluster member, its CMD position would align with sub-subgiant stars, which are typically associated with interacting binaries \citep{Geller_2017}.

\subsubsection{s04}
Two HST counterparts, s04a and s04b, fall within the 1$\sigma$ X-ray positional uncertainty of s04. s04a lies along the main sequence in the CMD, whereas s04b is unusually blue, occupying a region often associated with CVs. In the X-ray/optical flux ratio diagram, both sources fall between the AB and CV regimes.

\subsubsection{s11}
Two HST counterparts, s11a and s11b, lie within the 1$\sigma$ X-ray positional uncertainty. In the CMD, s11a is located in the blue straggler region, while s11b lies near the RGB or red clump. Both sources fall within the AB region in the X-ray/optical flux ratio diagram. An eclipsing binary system, OGLE-BLG-ECL-290997, is cataloged at this position with a period of $P\sim1.029$~d \citep{Soszynski2016}. Due to limited spatial resolution, OGLE photometry cannot distinguish between s11a and s11b.
\\
\\
\\
Overall, these supplementary counterpart candidates display CMD positions and X-ray properties that could be consistent with interacting binaries or other active systems, though their probable field star nature introduces uncertainties. Further observations could help confirm their nature.

\section{Summary \& Conclusion}

This paper has presented the first X-ray study of NGC 6528, one of the most metal-rich GCs in the Galaxy. We find a healthy collection of faint X-ray sources, all with luminosities in the range $L_X \sim 10^{31}$--$10^{32}$ erg s$^{-1}$. By combining X-ray data with optical observations primarily from the Hubble Space Telescope, we argue that the X-ray sources are primarily CVs and ABs, though one or more of the brighter sources could be quiescent low-mass X-ray binaries. This appears to be a cluster where the strong tidal effects of being close to the centre of the Galaxy have dominated its evolution, leading to extensive dynamical mass loss. Future time-series optical or near-IR photometry could improve the classification of the detected sources. As the cluster is detected in GeV $\gamma$-rays, the future discovery of MSPs is a distinct possibility, which would improve our understanding of the dynamical state and compact object populations in the cluster.

\section*{Acknowledgements}
We acknowledge support from NASA grant 80NSSC21K0628. Based on observations made with the NASA/ESA Hubble Space Telescope, and obtained from the Hubble Legacy Archive, which is a collaboration between the Space Telescope Science Institute (STScI/NASA), the Space Telescope European Coordinating Facility (ST-ECF/ESAC/ESA) and the Canadian Astronomy Data Centre (CADC/NRC/CSA). We thank Craig O. Heinke for helpful guidance during and after the review process, and Edoardo P. Lagioia for providing the membership catalog and related clarifications.

\section*{Data Availability}
The data in this paper can be accessed from \textit{Chandra} Data Archive (\url{https://cda.harvard.edu/chaser}). 
The HST data can be retrieved from the Mikulski Archive for Space Telescopes (MAST) Portal (\url{https://mast.stsci.edu/portal/Mashup/Clients/Mast/Portal.html})  with the proposal IDs listed in \autoref{tab:HSTdata}. 
This work has made use of data from the European Space Agency (ESA) mission
\textit{Gaia} (\url{https://www.cosmos.esa.int/Gaia}), processed by the \textit{Gaia} Data Processing and Analysis Consortium (DPAC, \url{https://www.cosmos.esa.int/web/gaia/dpac/consortium}). 



\bibliographystyle{mnras}
\bibliography{ref} 






\bsp	
\label{lastpage}
\end{document}